\documentclass[fleqn,10pt]{wlscirep}
\usepackage[utf8]{inputenc}
\usepackage[T1]{fontenc}
\usepackage{lineno}
\usepackage{layouts}

\title{Baryonic properties of nearby galaxies across the stellar-to-total dynamical mass relation}

\author[1,2*]{Laura Scholz-D\'iaz}
\author[1,2]{Ignacio Mart\'in-Navarro}
\author[1,2]{ Jes\'us Falc\'on-Barroso}
\author[3]{Mariya Lyubenova}
\author[4]{Glenn van de Ven}
\affil[1]{Instituto de Astrof\'isica de Canarias, La Laguna, 38205, Spain}
\affil[2]{Universidad de La Laguna, Departamento de  Astrof\'isica, La Laguna, 38200, Spain}
\affil[3]{European Southern Observatory, Garching bei München, 85748, Germany}
\affil[4]{University of Vienna, Department of Astrophysics, Vienna, 1180, Austria}
\affil[*]{scholz@iac.es}

\newcommand\aap{A\&A}                
\newcommand\aj{AJ}                   
\newcommand\apj{ApJ}                 
\newcommand\apjl{ApJ}                
\newcommand\apjs{ApJS}               
\newcommand\araa{ARA\&A}             
\newcommand\mnras{MNRAS}             
\newcommand\nat{Nature}              
\newcommand\pasp{PASP}               

\usepackage[superscript,biblabel,nomove]{cite}

\begin{abstract}
\textbf{In the standard cosmological model, the assembly of galaxies is primarily driven by the growth of their host dark matter halos. At the center of these halos, however, baryonic processes take over, leading to the plethora of observed galaxy properties. The coupling between baryonic and dark matter physics is central to our understanding of galaxies and yet, it remains a challenge for theoretical models and observations. Here, we demonstrate that measured ages, metallicities, stellar angular momentum, morphology and star formation rates, correlate with both stellar and halo  mass. Using dynamical modeling, we find that at fixed stellar mass, CALIFA galaxies become younger, more metal-poor and rotationally supported, have higher star formation rates and later-type morphologies as their total mass increases, with independent stellar and total masses measurements. These results indicate that the formation of galaxies and thus their baryonic properties do not vary with stellar mass alone, with halo mass also playing an important role.}
\end{abstract}

\begin{document}

\flushbottom
\maketitle

\thispagestyle{empty}

Historically, galaxies in the Local Universe have been studied with regards to luminous matter alone, being traditionally characterized in terms of fundamental baryonic properties such as stellar mass, size, kinematics, or chemical composition \cite{1976ApJ...204..668F,1977A&A....54..661T,1987ApJ...313...59D,2005ApJ...621..673T} to unravel the intricate processes taking part in their baryonic cycle (e.g, gas cooling, star formation, chemical enrichment, supernovae and black hole feedback). However, in our favored cosmological paradigm, the formation of galaxies is, to first order, primarily driven by the growth of their host  dark matter halos \cite{1978MNRAS.183..341W,1984Natur.311..517B}, with halo assembly being broadly understood theoretically due to large gravity-only N-body simulations \cite{2005Natur.435..629S}. Nevertheless, to understand the galaxy formation process in this cosmological context is crucial not only to understand how dark matter halos assemble and the complex baryonic physics occurring at the center of these halos, but also the interplay between the two. 

A common way to investigate the connection between galaxies and halos is through the stellar-to-halo mass relation (SHMR) for central galaxies \cite{2018ARA&A..56..435W}, which links galaxy stellar masses and halo masses, and whose scatter is related to the efficiency of galaxy formation. Although this scatter has been studied in terms of a variety of galaxy properties (e.g, morphology, color, star formation rate), results from different observational and theoretical techniques seem to disagree \cite{2018ARA&A..56..435W}. While some works find that more massive galaxies at a given halo mass have characteristics of old, red and passive systems  \cite{2016MNRAS.455..499L,2018MNRAS.477.1822M,2019ApJ...874..114C,2022MNRAS.511.4900S,2022ApJ...933...88O}, others suggest that they have young, blue, star-forming ones \cite{2011MNRAS.410..210M,2016MNRAS.457.3200M,2015ApJ...799..130R,2020MNRAS.499.3578C,2021NatAs...5.1069C}. At the same time, several studies suggest that the scatter also correlates with halo formation time or concentration using different methods \cite{2013MNRAS.431..600W,2017MNRAS.470.3720T,2017MNRAS.465.2381M,2021MNRAS.505.5117Z,2021NatAs...5.1069C}. Halo assembly bias is the dependence on halo clustering on a secondary property other than mass \cite{2005MNRAS.363L..66G,2018MNRAS.474.5143M}. When that halo property also has an impact on the SHMR relation (also known as occupancy variation\cite{2018ApJ...853...84Z, 2018MNRAS.480.3978A}), the halo assembly bias can propagate to galaxy clustering (i.e. galaxy assembly bias\cite{2007MNRAS.374.1303C}). Yet, the nature of the coupling between the baryonic cycle and halo assembly remains unclear, and whether, how and why observed galaxy properties are affected by halo properties is highly debated.

One of the main drawbacks for understanding how halo assembly affects the evolution of galaxies is that measuring properties of dark matter halos (such as mass or formation time) is still a challenge for current observational facilities, mainly due to the nature of dark matter itself. Historically, galaxy stellar and gas dynamics measured up to large distances have provided dynamical evidence for dark matter in the Local Universe through the study of galaxy rotation curves (the circular velocity of stars and gas vs. their galactocentric distance) \cite{1966ApJ...144..639R,1970ApJ...159..379R}, playing a key role in the discovery of dark matter. However, to study the galaxy-halo connection, observational works typically employ halo mass estimations based on alternative methods (e.g., weak gravitational lensing \cite{2016MNRAS.457.3200M}, satellite kinematics \cite{2011MNRAS.410..210M}, group and cluster catalogs \cite{2007ApJ...671..153Y}), with only a few recent works inferring halo masses dynamically (using globular clusters \cite{2021A&A...649A.119P} and HI cold-gas rotation curves \cite{2019A&A...626A..56P}).

Here, we take an observational approach based on stellar dynamics to investigate the interplay between baryons and dark matter within the galaxies. We assess the total dynamical mass of the galaxies as an alternative metric sensitive to their dark matter content. We introduce the stellar-to-total dynamical mass relation for nearby galaxies drawn from the Calar Alto Legacy Integral Field Area (CALIFA) integral-field spectroscopic (IFS) survey, and study the behavior of key baryonic galaxy properties across the relation. Total dynamical masses of galaxies are derived through detailed dynamical modelling of their stellar kinematics using IFS data either by solving the equations of stellar hydrodynamics \cite{1922MNRAS..82..122J} or by using the numerical orbital-superposition method \cite{1979ApJ...232..236S}. In this sense, these are dynamical tracers of the total mass enclosed within a given aperture, being sensitive to the motions and dynamics of the stars and gas within that aperture. Thus, these total masses contain information not only about the baryonic component, but also about the dark matter content of the galaxies, given that the dynamical modelling involved incorporates a dark matter component to be able to reproduce the stellar dynamics \cite{2006MNRAS.366.1126C}.

\section*{Stellar-to-total dynamical mass relation}

We used data of 260 galaxies from the CALIFA survey, which provides spatially resolved optical spectra for each galaxy. This set of nearby galaxies (in the redshift range $0.005 \leq z \leq 0.029$) were observed with a high spectral resolution to derive high-quality stellar kinematics, and are representative of the CALIFA mother sample \cite{2014A&A...569A...1W}, with morphologies ranging from ellipticals to late-type spirals. The sample is largely biased towards central (80\%, based on subsample\cite{2007ApJ...671..153Y}), isolated galaxies with no signs of interactions. For each galaxy, its stellar mass ($M_{\star}$), mean age and metallicity ([M/H]), apparent angular momentum ($\lambda_{Re}$), star formation rate (SFR) and morphology are measured (Methods). The total dynamical masses of the galaxies (enclosed within 3 half-light radii ($R_e$) are inferred by constructing axisymmetric dynamical Jeans models which fit the stellar kinematics (Methods).

\subsection*{Ages and metallicities}

By looking at the stellar population properties of our CALIFA galaxies across stellar-to-total dynamical mass relation (Fig.~\ref{fig:sdmr_stelpops}) we observe that galaxy ages (left panel) and metallicities (right panel) depend not only on stellar mass, but also on the total dynamical mass. At fixed total mass, galaxies become older and more metal-rich as stellar mass increases, specially for intermediate masses $ \rm \left( 10^{10} M_{\odot} \leq M_{\star} \leq 10^{11} M_{\odot} \right) $.  While at the same time, for a given stellar mass, galaxies with higher total masses are also younger and more metal-poor compared to those with lower total masses.

The dependence of these baryonic parameters on stellar mass and total dynamical mass is quantified through a partial correlation analysis (Methods), which allow to assess these dependencies by removing inter-correlations between the data. The bottom right corner of the panels of Fig. \ref{fig:sdmr_stelpops} shows the partial correlation coefficient strengths (solid black lines) between the stellar population parameters and stellar mass (vertical line) and total dynamical mass (horizontal line). These coefficients (which employ the Spearman rank correlations between the parameters of interest for their computation) indicate that there is a statistically significant correlation (anti-correlation) between the stellar population parameters and the stellar mass (total dynamical mass). To guide the eye, the direction of the maximal increase of the stellar population properties is also indicated as a red line (whose slope is computed using the partial correlation coefficients). Despite that the correlation with stellar mass is stronger for both age and metallicity, the direction of maximal increase is not vertical neither for age or metallicity, indicating that although their primary driver is stellar mass, total dynamical mass also has a secondary but noticeable role.

A key point of our findings is that there is an effect seen in stellar population properties of the galaxies, which can not be attributed to stellar mass alone, and can be captured by a dynamical tracer sensitive to the motions of the stars within the galaxies. These total masses are derived through detailed dynamical modelling, which need to incorporate dark matter component (in addition to the stellar one) to be able to fit the stellar dynamics. This hints that the baryonic cycle of galaxies is indeed modulated by the properties of the host dark matter halos.

\subsection*{Other baryonic galaxy properties}

Going further, we investigate whether there are other galaxy properties also sensitive to this dynamical tracer of total mass, as different baryonic observables can probe different characteristic time-scales of the galaxy formation process. We assess the behavior of SFRs (sensitive to the current level of star formation), the apparent angular momentum, $\lambda_{Re}$ (sensitive to the dynamics of the stars) and Hubble type (indicator of morphology) across the stellar-to-total dynamical mass relation (Fig. \ref{fig:sdmr_sfr_lambda_morph}). Along with the stellar population properties, $\lambda_{Re}$ (left panel), SFR (middle panel) and Hubble Type (right panel) also show dependencies with both stellar mass and total dynamical mass, which are specially noticeable for intermediate stellar masses. Analogously to Fig. \ref{fig:sdmr_stelpops}, these dependencies are also quantified through a partial correlation analysis (Methods). The scatter of the relation at fixed total mass anti-correlates with SFR, morphology and $\lambda_{Re}$, with more massive galaxies being less rotationally supported, having lower SFRs and earlier-type morphologies than less massive ones. But most interestingly, these baryonic properties also correlate with the total mass at fixed stellar mass, as galaxies with higher total masses are also more rotationally supported, have higher SFRs and later-type morphologies than galaxies with lower total masses. 

Connecting with our previous findings, not only stellar population properties are correlated with dynamical mass, but also other elements of the baryonic cycle related to the level of star formation, galaxy morphology and kinematics. Furthermore, the stellar and total mass dependencies of the stellar populations are consistent with the ones of these other baryonic properties in the sense that the scatter of the relation seems to be correlated with the evolutionary stage of the galaxies. At fixed total mass, more massive galaxies have characteristics of more evolved systems --older, more metal-rich and more dispersion dominated galaxies with lower SFRs and early-type morphologies--, while less massive galaxies of less evolved systems --younger, more metal-poor and more rotationally supported galaxies have higher SFRs and late-type morphologies--. Note that an analogous argument can be donein the reverse direction at fixed stellar mass. 

Moreover, this picture is also consistent with the recent findings of ref. \cite{2023MNRAS.521.4173B}, who show that the behavior of gas-phase metallicities across this relation (their Figure 2a) is very similar to the ones we found for all these galaxy observables.  These authors found that gas-phase metallicity does not only exhibit a primary dependence on stellar mass, but most interestingly, also an anti-correlation with total mass (also quantified through partial correlation analysis). We note that the dependence they found is weaker than ours, but in this case they study spiral galaxies only, and the total dynamical masses are estimated up to only 1 $R_e$. It is specially noteworthy that stellar and gas-phase metallicities are both tracers of the metallicity of the (star-forming) gas, but at very different epochs of the galaxy formation process, and they still show these very similar dependencies.

\section*{Discussion}

\begin{figure*}
    \centering
    \includegraphics[width=\linewidth]{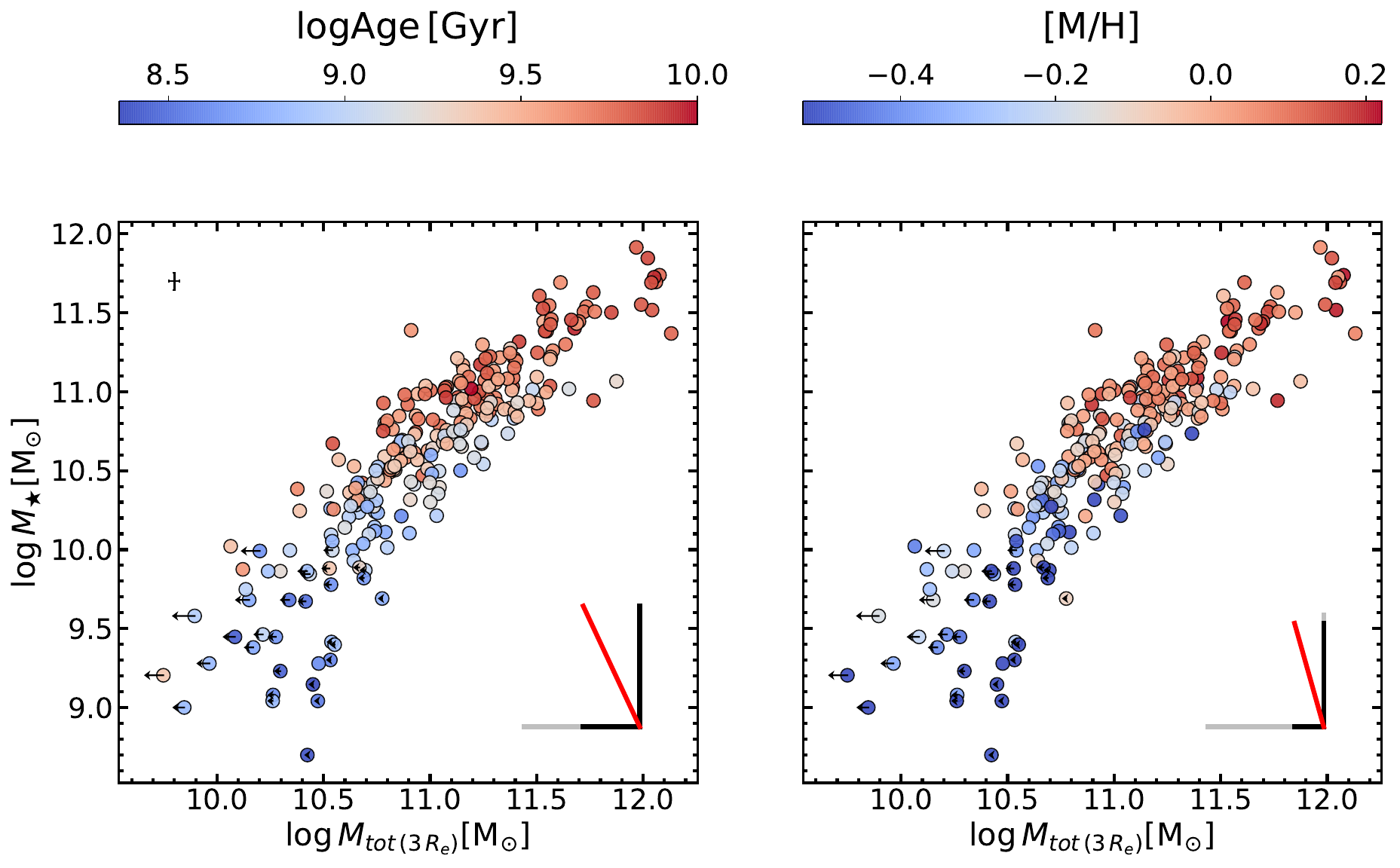}
    \caption{Stellar-to-total dynamical mass relation for CALIFA in terms of galaxy stellar population properties. Galaxies are shown as circles colored-coded by mean galaxy ages (left panel) and metallicities (right panel). 1$ \rm \sigma$ errors are indicated in the upper left corner (left panel). Partial correlation coefficient strengths (Methods) are shown in the bottom right corner (solid black lines) between the stellar population parameters and $M_{\star}$ (vertical) and $M_h$ (horizontal). Grey solid lines have a length which corresponds to a correlation coefficient of 0.6 for reference. The direction of maximal increase of the stellar population parameters (Methods) is indicated as a red solid line. The arrows show a correction to account for the gas contribution to the total mass (stars + dark matter) of low-mass late-types.}  
    \label{fig:sdmr_stelpops}
\end{figure*}

\begin{figure*}
    \centering
    \includegraphics[width=\linewidth]{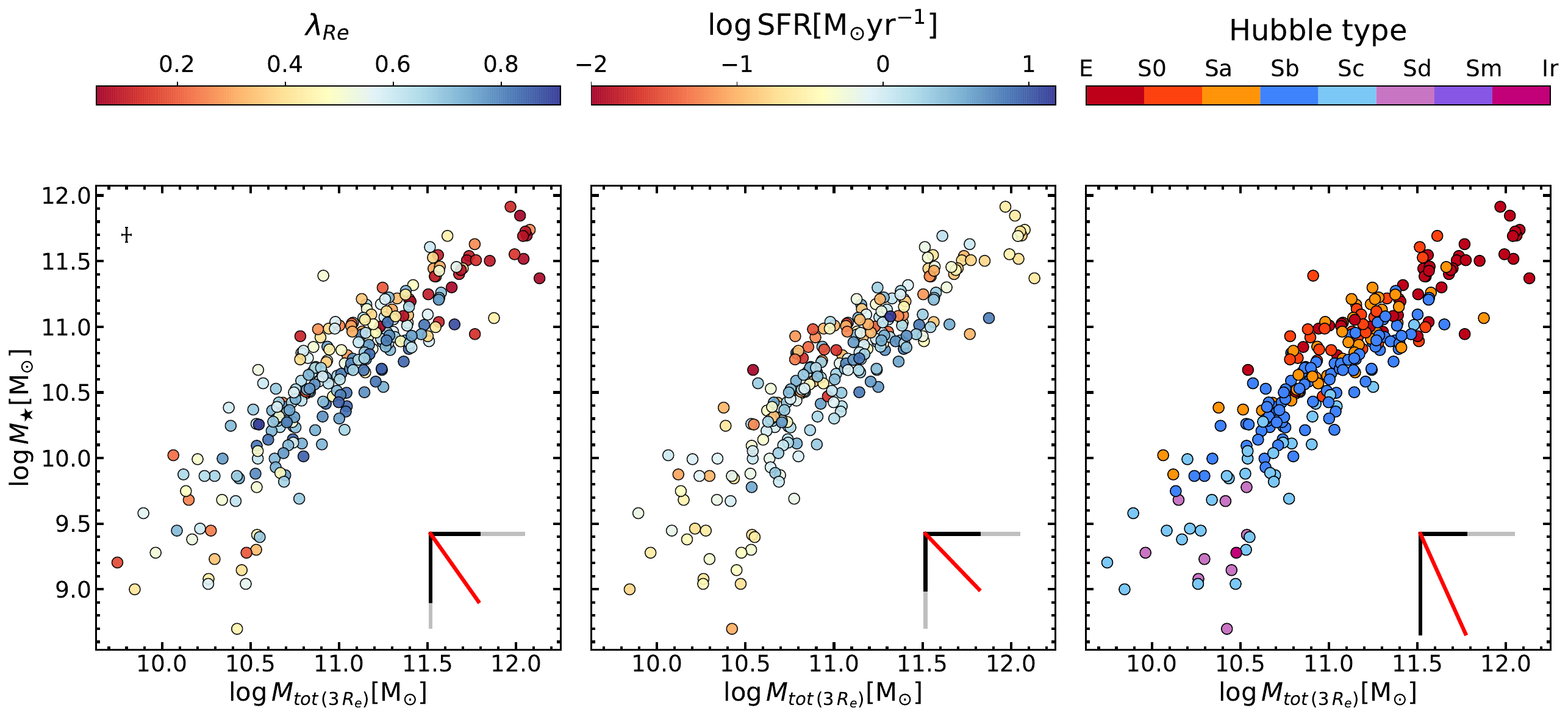}
    \caption{Stellar-to-total dynamical mass relation for CALIFA in terms of different baryonic properties. Galaxies are shown as circles colored-coded by the apparent angular momentum ($\lambda_{Re}$) (left panel), galaxy star formation rate (middle panel) and Hubble type (right panel). 1$ \rm \sigma$ errors are indicated in the upper left corner (left panel). Partial correlation coefficient strengths (Methods) are shown in the bottom right corner (solid black lines) between the different galaxy properties and $M_{\star}$ (vertical) and $M_h$ (horizontal). Grey solid lines have a length which corresponds to a correlation coeffcient of 0.6 for reference. The direction of maximal increase of the stellar population parameters (Methods) is indicated as a red solid line.} 
    \label{fig:sdmr_sfr_lambda_morph}
\end{figure*}

We find direct evidence indicating that different observables of the baryonic cycle are connected to the scatter of the stellar-to-total dynamical mass relation.  We find that mean ages, metallicities, global star formation rates, morphologies and the apparent momentum of the galaxies depend both on stellar mass and total mass (enclosed within 3 $R_e$ derived through Jeans dynamical modelling). At fixed total mass, galaxies become older and more metal-rich, have lower SFRs and are less rotationally supported with increasing stellar mass (Fig. \ref{fig:sdmr_stelpops}) and (Fig. \ref{fig:sdmr_sfr_lambda_morph}).

Figs. \ref{fig:sdmr_stelpops} and \ref{fig:sdmr_sfr_lambda_morph} seem to suggest that the baryonic cycle of galaxies might be modulated by halo properties across cosmic time, given that stellar mass is not the only driver of these different galaxy properties, as total dynamical mass also has a secondary role.  It is worth noting that, unlike previous studies, our stellar and total masses are determined in a completely independent manner. While both quantities are sensitive to the stellar content of the galaxies, the total mass measurements are a dynamical tracer sensitive to the stellar motions, and hence, provide additional information about the dark matter content within the galaxies (which is not captured by stellar mass). Moreover, the case of the stellar populations is very interesting, given that a similar behavior is found for galaxy ages and metallicities across the stellar-to-halo mass relation for SDSS central galaxies for low/intermediate mass halos \cite{2022MNRAS.511.4900S}, as the scatter of the SHMR correlates with galaxy ages and [M/H] (at fixed halo mass). Moreover, ages and metallicities of SDSS galaxies depend primarily on stellar mass and secondarily on halo mass, while in the case of CALIFA the main driver of these properties is again stellar mass, although with a secondary dependence on total mass. We also note the direction these properties increase (quantified through partial correlations) increase qualitatively following a similar direction in both relations. In this sense, ages and metallicities show a very similar behavior across the SHMR and the stellar-to-dynamical mass relation, with the total dynamical masses (enclosed within 3 $R_e$) of the galaxies mimicking the behavior of the mass of their host halos. 

\subsection*{Connection with the scatter of the SHMR}

Under the assumption that total dynamical mass indeed traces the mass of the host halos of the galaxies, the scatter of the stellar-to-dynamical mass relation could give insights into the physical origin of the scatter of the SHMR. Given the similarities we found between stellar population properties across the stellar-to-total dynamical mass relation and the SHMR, the fact that different baryonic galaxy properties indicate that the scatter of the stellar-to-total dynamical mass relation correlates with galaxy evolutionary stages can be interpreted as a signature of the time in which dark matter halos assemble. Many theoretical works based on different techniques predict that the scatter of the SHMR correlates with halo formation time \cite{2013MNRAS.431..600W,2017MNRAS.470.3720T,2017MNRAS.465.2381M,2018ApJ...853...84Z,2021NatAs...5.1069C}, generally indicating that more massive galaxies at fixed halo mass reside in earlier formed halos. This could naturally explain our findings, with earlier-formed halos hosting galaxies that assembled the bulk of their stars early on and quenched also earlier, ending up with older and more-metal rich stellar populations at present day, low SFRs, and being more dispersion-dominated with typically early-type morphologies. However, it is important to note that this connection between halo formation time and the scatter of the SHMR is strongly dependent on the simulation and/or theoretical model used (see section 5.2.2 in ref.\cite{2022ApJ...933...88O}). As an example, while the stochasticity of late mergers can dim this dependence on halo assembly time at high halo masses in hydro-simulations \cite{2017MNRAS.465.2381M}, this connection is strongest for the high halo mass regime in semi-analytics models \cite{2019MNRAS.488.3143B}, as it is driven by accreted stellar populations coming from mergers \cite{2020MNRAS.493..337B}. We also note that in our observational approach we measure baryonic properties within one half-light radii, and hence we are probing the central regions of the galaxies, which are dominated by stellar populations formed in-situ. In a forthcoming paper we are analyzing stellar population gradients and radial dependence of SFHs, which could also bring light into this matter. On the other hand, state-of-the art zoom-in hydro-simulations also predict that halo assembly time can influence the evolution of the star formation histories of galaxies\cite{2021MNRAS.501..236D}.  These authors find that modifying halo assembly histories, essentially shifting them earlier or later at fixed halo mass, can affect present-day galaxy properties. These simulations predict that an earlier-formed halo host a quenched galaxy, which has suffered a decline in rotational support and a strong morphological evolution from disc-like to spheroidal, and a lower gas fraction, while one hosted by later-formed halos remains star-forming, disk-like and gas-rich. In these simulations these differences are essentially driven by the combination of halo assembly and black hole feedback, as earlier halos foster the growth of more massive central black holes, which inject more feedback energy into the halos, effectively quenching the galaxies. However, we caution the reader that the coupling between galaxy and halo properties is heavily sensitive to the baryonic recipes included in the subgrid-scale physics of the simulations, an issue which we also discuss below.

Our findings suggest that investigating the scatter of the stellar-to-dynamical mass relation could help to understand the connection between the baryonic cycle and the scatter of SHMR. Despite the fact that the scatter of the SHMR has been widely studied in terms of different galaxy properties, results from different methods and techniques do not portray a clear picture \cite{2018ARA&A..56..435W}. While some works find that more massive galaxies (at fixed $M_h$) have redder colors \cite{2016MNRAS.455..499L}, older stellar populations \cite{2022MNRAS.511.4900S,2022ApJ...933...88O} or are passive \cite{2016MNRAS.455..499L,2018MNRAS.477.1822M,2019ApJ...874..114C}, other findings indicate that more massive halos (at fixed $M_{\star})$ host redder \cite{2011MNRAS.410..210M,2016MNRAS.457.3200M,2021NatAs...5.1069C} and older galaxies \cite{2022ApJ...933...88O}, suggesting the opposite trend, with some authors even finding that more massive galaxies have bluer colors \cite{2015ApJ...799..130R} and are more rotationally supported \cite{2020MNRAS.499.3578C} (at fixed $M_h$). 
In principle, some of these results seem to be contradictory, but they come from a variety of different analyses and techniques (e.g., numerical simulations, empirical models and observational works employing different halo mass estimations). While accurately determining halo masses to study the SHMR is still an observational challenge, the case for theoretical models is also not trivial. Understanding how the complex baryonic processes (e.g., supernovae and black hole feedback, star formation, chemical enrichment) 
occurring at small scales within the galaxies (e.g., smaller than the resolution limit of current state-of-the art cosmological hydro-simulations) are coupled to halo assembly is still highly debated, with this coupling  depending on the implementation of the baryonic physics within the models \cite{2020MNRAS.499.3578C,2021NatAs...5.1069C,2021MNRAS.501..236D,2020MNRAS.491.4462D}. Bringing light to this matter, our findings seem to be in line with works that find higher stellar mass galaxies at a given halo mass having characteristics of old, red and passive systems, while also calling for a revision of the implementation of baryonic processes in numerical simulations, which also need to take into account how the growth of galaxies and halos are connected in order to reproduce observed galaxy properties. 

Furthermore, note that in this work we follow ref. \cite{2022MNRAS.511.4900S} and analyze baryonic properties across the 2D stellar-to-total dynamical mass plane, instead of 1D projections of it, given the complex trends of the stellar population properties across the SHMR and the so-called `inversion problem'. It has been noted that analyzing the plane $M_h$ vs. $M_{\star}$ instead of the $M_{\star}$ vs. $M_h$ plane can affect or even reverse the resulting trends when galaxies are divided into two distinct populations (e.g., passive/star-forming or red/blue), and at the same time, they are binned in terms of the x-axis of the relation ($M_h$ or $M_{\star}$) \cite{2018MNRAS.477.1822M,2020MNRAS.499.4748M,2020MNRAS.499.3578C}. Thus, binning in $M_h$ or $M_{\star}$ may lead to contradictory trends given the different statistics of the galaxy populations (e.g. star-forming or passive) at fixed stellar or halo mass, making this problem an important source of discrepancies between the different works mentioned above studying the SHMR. Moreover, at first, the results of ref. \cite{2020MNRAS.499.4748M} studying the  $\rm M_{\star}- M_h$ relation and ref.  \cite{2016MNRAS.457.3200M}, who in turn analyzed the $\rm M_h-M_{\star}$ relation, seem to suggest opposite trends. Nevertheless, ref. \cite{2020MNRAS.499.4748M} shows that they recover trends in agreement with ref. \cite{2016MNRAS.457.3200M} when using similar stellar mass bins in the $\rm M_h-M_{\star}$ plane.  

\subsection*{Total dynamical mass vs. Halo mass}
The hypothesis that total dynamical mass traces the halo mass of the galaxies, is subject to known caveats and limitations. Theoretically, virial halo masses can be defined as a function of overdensity (e.g., $M_{200}$, 
$M_{500}$). $M_{X}$ is defined as the mass enclosed within a sphere of radius $R_{X}$ (the radius of a sphere containing an overdensity X times the critical density of the Universe), where 200 or 500 are common used values for X. 
 However, these quantities are currently inaccessible to observations, given that in order to obtain halo masses directly, the measurements need to be extended up to large distances. Our total dynamical masses for CALIFA galaxies are limited by the integral-field unit spatial coverage, and they cover up to 3 half-light radius. Yet, they are an alternative observational method that can trace the dark matter content of nearby galaxies \cite{2013MNRAS.432.1709C,2022FrASS...8..197N} and a first step towards directly measuring halo masses. Different methods suggest that although galaxies are not typically dominated by dark matter in the inner regions (e.g.  1 half-light radius), they tend to have non-negligible dark matter fractions \cite{2010ApJ...724..511A,2011MNRAS.415..545T,2013MNRAS.432.1709C,2018MNRAS.481.1950L,2022FrASS...8..197N}. However, note that we are not interested in the absolute amount of dark matter within the inner regions, but in the fact that, even with a low dark matter fraction in those regions, the total dynamical mass seems to map the virial halo mass after accounting for the stellar mass dependence. To directly asses our underlying assumption that total dynamical mass traces  halo mass, we have compared our dynamical masses to halo mass estimates from a group/cluster catalog based on abundance-matching \cite{2007ApJ...671..153Y}, finding that at fixed stellar mass, galaxies with higher dynamical masses are also assigned to higher-mass halos (Methods). Such an agreement between two independent halo observables reinforces our interpretation. Furthermore, we have also used the large cosmological numerical hydro-simulation from the EAGLE suite \cite{2015MNRAS.446..521S} to demonstrate that the total masses of galaxies at the typical aperture probed by the CALIFA data map halo masses, $M_{200}$ (at fixed stellar mass) (Methods). The correlation between these two quantities is remarkably  similar to that resulting from comparing CALIFA dynamical masses and  halo mass estimates based on group/cluster catalogs, cementing the reliability of our assumptions.

\subsection*{The role of different baryonic effects}
While our interpretation is based on the assumption that dynamical masses map the behavior of halo masses, baryonic processes can also contribute to the observed scatter in the stellar-to-total dynamical mass relation. In particular, aperture effects due to the morphology dependence of the effective radii could in principle partially mimic some of the observed trends. We have tested however that this aperture bias is not strong enough to generate the observed signal. Moreover, the total dynamical masses of late-type low mass galaxies (Sc/Sd in \ref{fig:sdmr_sfr_lambda_morph}) could be overestimated, given that they are expected to have a larger gas component contributing to the total mass compared to more massive early-types, i.e., a higher baryonic mass, which implies a lower contribution of dark matter to the total mass budget. For that, we estimate the gas contribution of these late-types using the equation shown in Fig. 19 of ref.\cite{2012ApJ...759..138P} and apply a correction to our total masses which is shown with arrows in Fig. \ref{fig:sdmr_stelpops}. We observe how this effect does not affect our conclusions, given that the correlation of the stellar-to-total dynamical mass relation and different galaxy properties is not clearly seen in this low-mass regime, and effectively the correction only reduces the total masses  (stellar + dark matter mass) slightly. Furthermore, one could argue that our findings are driven by systematic effects inherent to our dynamical models. However, it has been shown that derived quantities related to the total mass distribution (e.g., total enclosed masses) are reliable and seem to be insensitive to different model assumptions (see Methods). Additionally, we have tested that we recover our observed trends when using different total mass measurements (derived through Schwarzschild dynamical modeling) \cite{2018NatAs...2..233Z}. Finally, it is worth considering the effect that a variable stellar initial mass function (IMF) could have on the observed trends, as a change in the slope of the IMF has a non-negligible impact on the mass-to-light ratio (M/L) and hence on the derived stellar  masses. Directly deriving the IMF slope for our CALIFA galaxies using absorption spectra is unfeasible mainly due to the complex star formation histories of late-type galaxies, and because the predicted M/L would be heavily dependent on the functional form assumed for the IMF, as the M/L is dominated by very low-mass stars and stellar remnants \cite{2012Natur.484..485C, 2016MNRAS.463.3220L}. However, we can approximate the effect that a variable IMF would have on our sample by using the empirical relation between metallicity and IMF slope \cite{2015ApJ...806L..31M}. Note that this relation has only been established for quiescent early-types, but it is a reasonable assumption given our current understanding of IMF variations. We find that, at fixed total mass, galaxies with higher stellar masses have higher M/L as a consequence of their higher metallicities, older ages, and expected bottom-heavier IMF slopes. This would in fact suggest that the stellar mass in these objects is underestimated using a Milky Way-like IMF and thus, that the actual scatter in the stellar-to-total mass relation is larger than the one found in this work. Hence, in the context of a variable IMF, our findings should be then understood as a lower-limit on the actual dependence of the stellar population properties on total mass. \\

All in all, our results suggest a connection between the baryonic content and properties of galaxies and the mass of their host dark matter halos, which we hypothesize to be driven by secondary halo properties. Upcoming observational surveys that measure cold-gas up to large galactocentric distances, while also providing spectroscopic coverage of the baryonic component of the galaxies (e.g., WEAVE-Apertif) will be crucial to test this scenario, as they will allow to directly measure enclosed masses up to larger distances than CALIFA. In addition to that, observational estimations of additional halo properties, such as halo formation time, individually for large galaxy samples would be key to better understand the coupling between the baryonic cycle of galaxies and dark matter halo assembly.

\section*{Methods}
\subsection*{Galaxy sample and ancillary properties}
We use 260 galaxies with integral-field spectroscopic data from the CALIFA survey. These set of galaxies are representative of the full CALIFA mother sample \cite{2014A&A...569A...1W}, which was diameter-selected (45''< r-band angular isophotal diamater < 80'') from the Sloan Digital Sky Survey (SDSS) DR7 in the redshift range $0.005 \leq z \leq 0.03$. They were observed  with a high resolution configuration (R$\sim$1650 at $\sim$4500 \AA) over the spectral range 3650-4840~\AA \ in order to derive high-quality stellar kinematics, and also with lower resolution (R$\sim$850 at $\sim$5000 \AA) over a wider spectral range (3745-7300~\AA). Note that this galaxy sample is heavily biased towards central galaxies, with 95\% being isolated. 

These set of nearby galaxies have ancillary data available for the following galaxy properties: (i) the stellar masses are obtained from Sunrise spectral energy distribution fits \cite{2014A&A...569A...1W} assuming a Chabrier IMF; (ii) stellar population properties are computed by averaging light-weighted ages and mass-weighted [M/H] within 1~$R_e$ \cite{2015A&A...581A.103G}, using the full-spectral fitting code STARLIGHT \cite{2005MNRAS.358..363C} fed with a combination of the MILES models \cite{2010MNRAS.404.1639V} for SSPs older than 63 Myrs and the GRANADA \cite{2005MNRAS.357..945G} models for younger ages, adopting a Salpeter IMF; (iii) the apparent angular momentum, $\lambda_{Re}$ is drawn from ref. \cite{2019A&A...632A..59F}; (iv) present-day star formation rates (SFR) are based on Balmer-decrement corrected H$\alpha$ fluxes \cite{2017MNRAS.469.2121S}; and (v) the morphological classification is made by human by-eye \cite{2014A&A...569A...1W}. Regarding the morphological type, we have used the classification from ref. \cite{2014A&A...569A...1W} to classify the galaxies into eight different Hubble types following ref. \cite{2019A&A...632A..59F}, which are used in Fig. \ref{fig:sdmr_sfr_lambda_morph}. However, note that for the partial correlation analysis we directly used the classification from ref. \cite{2014A&A...569A...1W} .

\begin{figure*}
    \centering
    \includegraphics[width=0.9\linewidth]{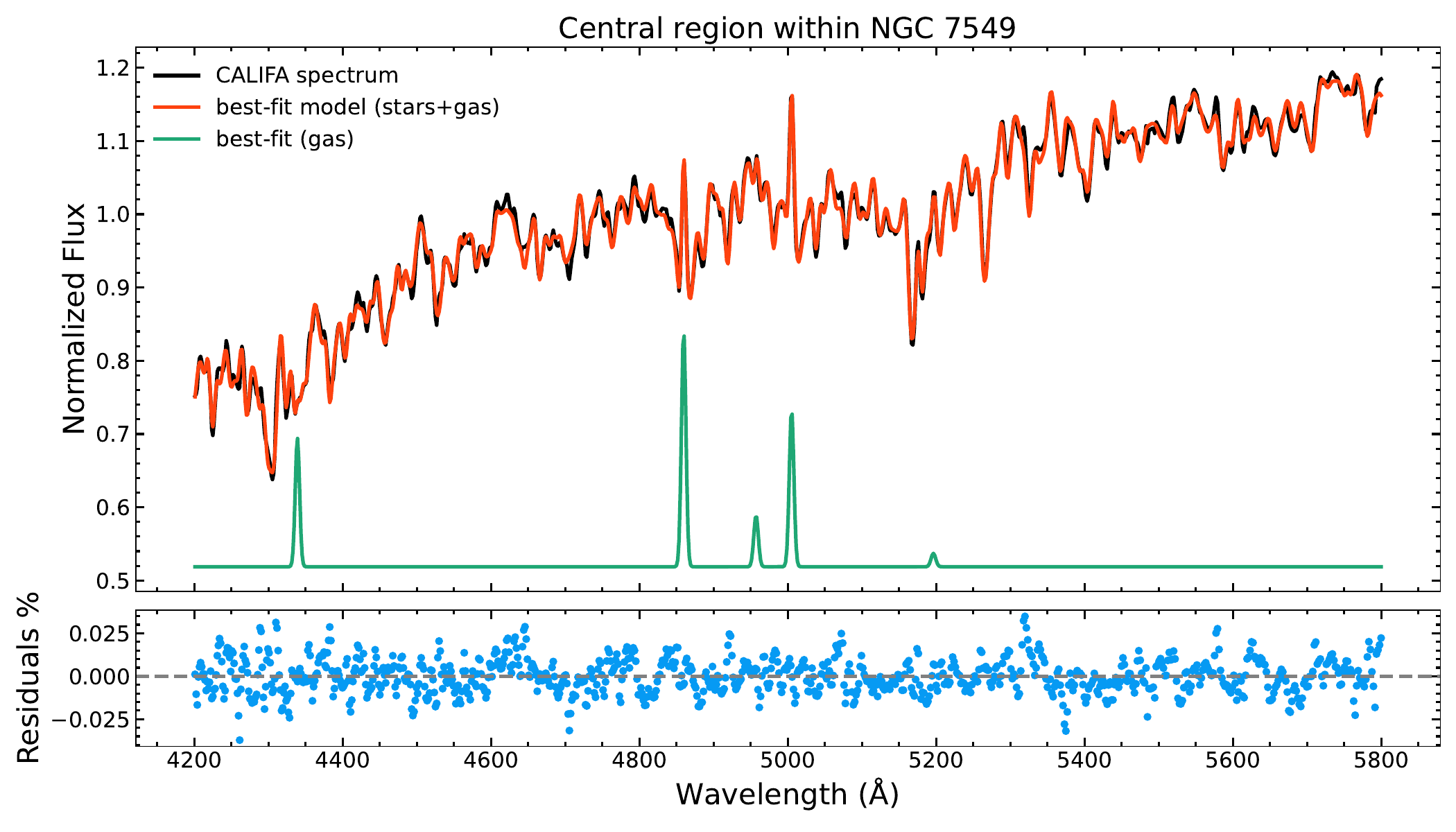}
    \caption{Spectral fit of the central region from NGC7549. The upper panel shows the galaxy spectrum in black, the best-fit model (stars and gas) in red, and the best-fit model of the gas in green. The residuals of the fit are shown in the bottom panel in blue.}
    \label{fig:fit}
\end{figure*}

The spectral fitting analysis was performed and described in detail in ref. \cite{2015A&A...581A.103G}. However, to illustrate the data, Fig. \ref{fig:fit} shows a fit to the spectrum of the central region from NGC7549, a typical galaxy in our sample. This fit was performed with the full-spectral fitting algorithm pPXF \cite{2004PASP..116..138C} fed with the MILES SSP models. The best-fit model corresponds to a linear combination of SSP models (stellar templates) and gas templates for nebular emission (modelled as gaussians) that best reproduce the galaxy spectrum (see ref. \cite{2022MNRAS.511.4900S} for more details on the method).

\subsection*{Total dynamical masses}
We derived the total dynamical masses of the galaxies by fitting axisymmetric dynamical models to their stellar kinematic maps. Our dynamical models are described in detail in a forthcoming paper (Lyubenova et al., {\it in prep}) therefore, we give here descriptive summary of the steps taken. We first parametrised the galaxies' stellar surface brightness by applying the multi-Gaussian expansion method (MGE) \cite{1994A&A...285..723E} to their $r$-band images from the $7^{th}$ data release of the Sloan Digital Sky Survey (SDSS)\cite{2009ApJS..182..543A}. We used the sky level estimates, obtained during the growth curve analysis of the CALIFA sample \cite{2014A&A...569A...1W}, to remove the sky contribution. We set a threshold of $\sim$1.5$\times$ the average standard deviation of the sky residual counts. We kept the position angle and the ellipticity fixed which are listed in Table~1 of ref. \cite{2017A&A...597A..48F}. 

Then we fitted the CALIFA stellar mean velocity and velocity dispersion fields \cite{2017A&A...597A..48F} with axisymmetric dynamical models, based on a solution of the Jeans equations as implemented by ref. \cite{2008MNRAS.390...71C}.  The fits were done over the full extent of the stellar kinematics maps up to within an elliptical aperture of 3~$R_e$\/. During the fits we allowed the velocity anisotropy in the meridional plane $\beta_z$ to vary in the range  $(-0.5,1)$, as well as the galaxy inclination. As a start we estimated the inclinations of the galaxies based on their global ellipticity \cite{2021ApJ...914...45V} and then allowed this value to vary within $\pm 20\deg$. The best-fit based on  $\chi^2$  statistics then yielded our dynamical mass-to-light ratios, which we converted into total dynamical masses by multiplying to the projected luminosity within elliptical aperture of 3~$R_e$\/. While in principle both the velocity anisotropy and stellar mass-to-light ratio may vary with radius, we found that this did not significantly improve our fits.

Our Jeans models are the so-called mass-follows-light (MFL) models, as their total mass distribution can include both a luminous and a dark matter component, with their combined total mass following the observed light distribution. However, note that it has been shown that derived quantities related to the total mass distributions obtained with good quality data are reliable against different mass model assumptions (e.g. mass-follows-light, free Navarro-Frenk-White (NFW) dark matter halo, fixed NFW, and generalized NFW)\cite{2013MNRAS.432.1709C,2023MNRAS.522.6326Z}.

\subsection*{Partial correlation analysis}
We quantify correlations between different galaxy properties and stellar/total mass while removing inter-correlations between the data by using partial correlation statistics.  This method is based on Spearman rank correlation, and the partial correlation coefficient between two variables, keeping a third parameter fixed \cite{2017MNRAS.471.2687B,2020MNRAS.492...96B} is defined as:

\begin{equation}
    \rho_{AB,C} = \frac{\rho_{AB} - \rho_{AC} \cdot \rho_{BC}}{\sqrt{1-\rho_{AB}^2}\sqrt{1-\rho_{BC}^2}}
    \label{eq:par_corr}
\end{equation}

which measures the correlation strength between variables A and B (at fixed C), with $\rho_{ij}$ being the Spearman rank correlation coefficient between parameters i and j. 

The direction of maximal increase of galaxy properties across the stellar-to-total dynamical relation is computed following \cite{2020MNRAS.492...96B}. The slope of the axis is given by the ratio of the partial correlation coefficient strengths $\rho_{BC,A}$ and $\rho_{AC,B}$, where A corresponds to $M_{tot}$, C corresponds to a given galaxy property (age or [M/H]), and B to $M_{\star}$.

\subsection*{Comparison with halo masses}

\begin{figure*}
    \centering
    \includegraphics[width=0.7\linewidth]{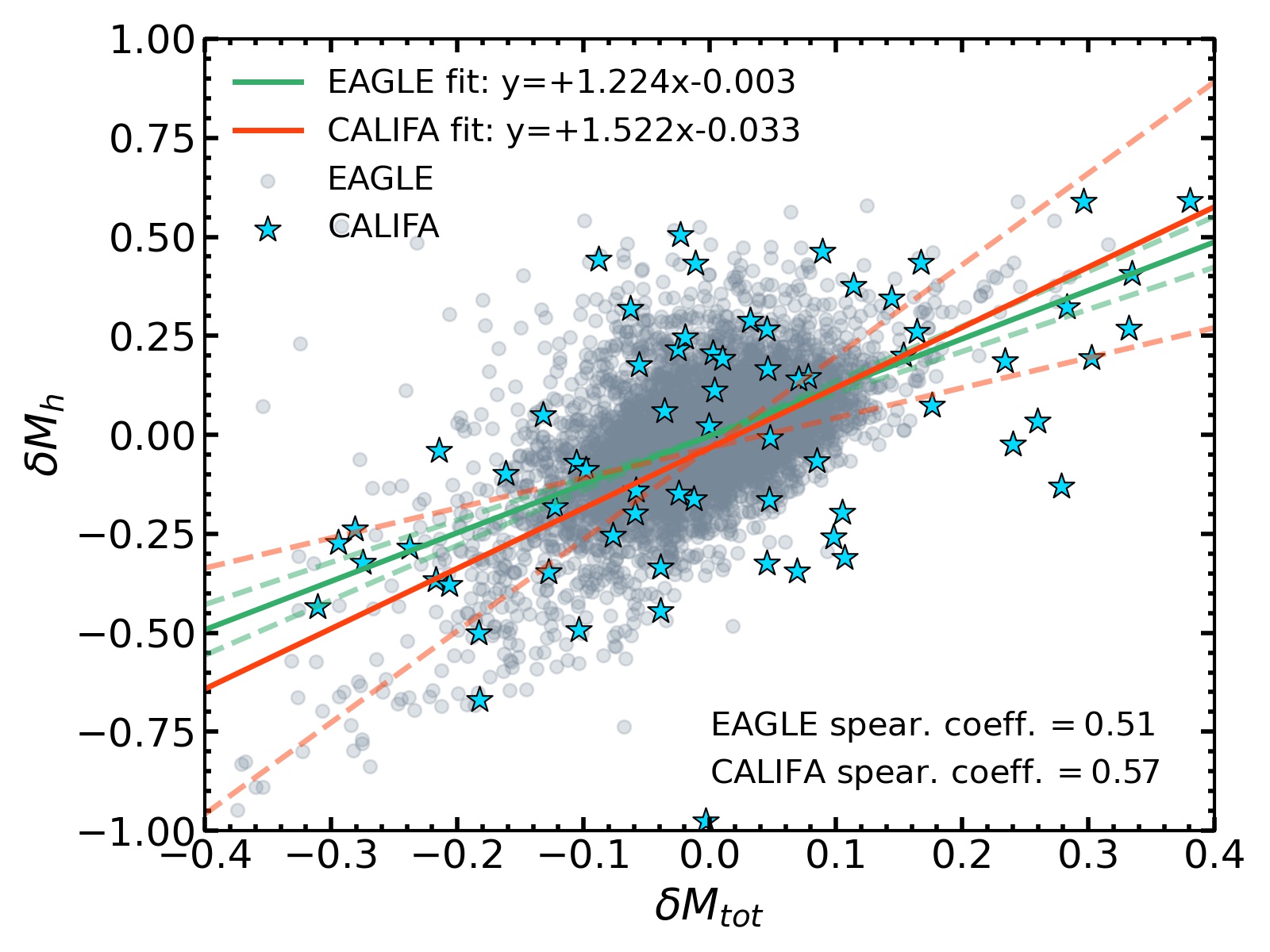}
    \caption{Total dynamical masses vs. halo masses. We show the residuals of the $\rm \log M_h - \log M_{\star}$ vs. $\rm \log M_{tot} - \log M_{\star}$ relations, i.e., $\delta M_h$ vs. $\delta M_{tot}$. CALIFA (EAGLE) galaxies are shown with blue (grey) stars (circles). 
    The solid lines correspond to the best-fitting relations and the dashed lines to the fit uncertainties for CALIFA and EAGLE, which are shown in orange and green, respectively. We also show the Spearman rank correlation coefficients. See the text for the details on the total and halo mass estimates.}
    \label{fig:residuals}
\end{figure*}

 We have compared our total dynamical masses with halo mass estimates from a group and cluster catalog. For that, we have cross-matched the central galaxies from the ref. \cite{2007ApJ...671..153Y} catalog with our galaxies, finding 70 objects in our CALIFA sample with abundance matching-based halo mass estimates. We followed ref. \cite{2022MNRAS.511.4900S} and assigned the halo mass of the group/cluster to its central galaxy, with the halo mass being computed using the total characteristic luminosity of the group/cluster.Although the galaxy sample is considerably reduced, we recover a very strong correlation between halo mass and total dynamical mass, with a Spearman correlation coefficient of 0.87. More relevant for the discussion of this work, we also removed the dependence on stellar mass by looking at the residuals of the $\rm \log M_h - \log M_{\star}$ vs. $\rm \log M_{tot (3 R_e)} -  \log M_{\star}$ relations (described in detail in section 6.1.1 of ref. \cite{2022MNRAS.511.4900S}). In Figure \ref{fig:residuals} we show how, at fixed stellar mass, galaxies with higher halo mass estimates also tend to have relatively higher total dynamical mass.

On the other hand, we also investigate whether total masses trace the halo masses in numerical simulations, where both halo and total masses are trivially derived. We use the publicly available database of the EAGLE suite of cosmological hydro-simulations \cite{2015MNRAS.446..521S,2016A&C....15...72M} (Methods). We selected central galaxies with $M_{\star} \geq 10^9$ at redshift zero from their 100 Mpc-box of the reference model, comparing halo masses ($M_h=M_{200}$) and total masses computed within an aperture of 20 kpc, given that the median 3 $R_e$ aperture of our galaxy sample is 17 kpc. Halo and stellar masses are provided in the database. Total masses correspond to the sum of the dark matter, gas, stellar and black hole total masses within 20 kpc. We find that halo masses tightly and strongly correlate with total masses with a Spearman correlation coefficient of 0.95, as expected. After removing the dependence on stellar mass as described above (i.e., exploring the residuals of the $\rm \log M_h - \log M_{\star}$ vs. $\rm \log M_{tot (20 \ kpc)} -  \log M_{\star}$ relations) we also find how in EAGLE, at fixed stellar mass, galaxies with higher total mass within 20 kpc are also those with relatively higher $M_{200}$  halo masses (Fig. \ref{fig:residuals}). This correlation predicted by simulations, quantified with a linear fit (green line), is very similar to the one found for CALIFA (orange line). This agreement between simulations and observations is remarkable, given that the quantities are derived in a completely independent manner. For observations we employ halo mass estimations from group/cluster catalogs, and in simulations they correspond to $M_{200}$. Moreover, our total masses are derived dynamically, while the total masses of the simulations come from the sum of all corresponding particles within a given aperture. We also note that it is beyond the scope of this work to assess the ability of cosmological simulations to reproduce total dynamical mass measurements, as we are interested in the fact that total mass within an aperture of 20 kpc traces $M_{200}$.

\begin{figure*}
    \centering
    \includegraphics[width=0.55\linewidth]{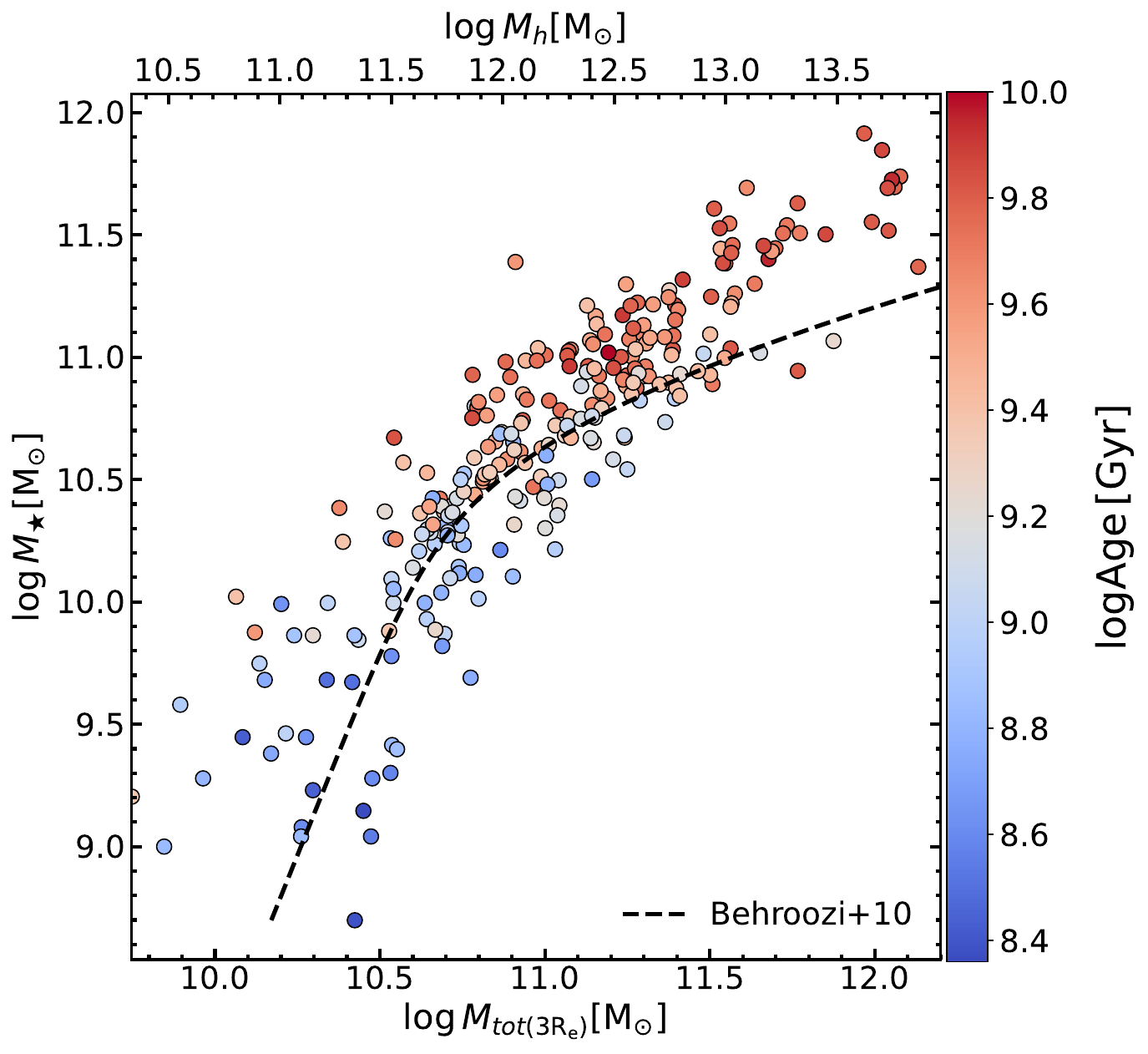}
    \caption{Analogous to the left panel of Figure 1, but in this case we show the halo mass scale in the upper horizontal axis. Halo masses are estimated using the linear fit to the correlation we found between our total dynamical masses and the halo masses from the group and cluster catalog ref. \cite{2007ApJ...671..153Y} ($M_{h(Yang+07)} =  +1.480 \times M_{tot(3 R_e)} - 4.092$) (based on the subsample in common). The black dashed line correspond to the equivalence of eq. 21 from ref. \cite{2010ApJ...717..379B} in this plane. }
    \label{fig:sdmr_beh}
\end{figure*}

Finally, we also illustrate the correspondence between our total dynamical masses and halo masses in Figure \ref{fig:sdmr_beh}. Based on the correlation we found between total dynamical mass and halo masses from the group and cluster catalog from ref. \cite{2007ApJ...671..153Y} (based on the subsample), we use the corresponding linear fit, $M_{h(Yang+07)} =  +1.480 \times M_{tot(3 R_e)} - 4.092$, to convert total masses into halo masses. Figure \ref{fig:sdmr_beh} is analogous to the left panel of Figure 1, but in this case we show the halo mass scale in the upper  horizontal axis. Additionally, to guide the eye we also show the equivalence of the mean SHMR from eq. 21 of ref. \cite{2010ApJ...717..379B} on the stellar-to-total dynamical mass plane using that conversion.

\subsection*{EAGLE simulation}

The \textsc{eagle} simulations \cite{2015MNRAS.446..521S} employ a modified version of the N-body Tree-Particle-Mesh smoothed particle hydrodynamics code \textsc{gadget-3} \cite{2005MNRAS.364.1105S} with subgrid recipes for radiative cooling, star formation, stellar evolution, chemical enrichment, stellar and black hole feedback (see ref. \cite{2015MNRAS.446..521S} for more details). The simulations assume a flat $\Lambda$CDM cosmology with parameters derived from Planck data \cite{2014A&A...571A...1P} and track the evolution of baryonic (stars, gas and black holes) and non-baryonic matter (dark matter) from redshift 127 to zero. The \textsc{eagle} reference model is performed in a 100 Mpc-box with initially the same number of dark matter and gas particles ($1504^3$). The initial dark matter and gas particle masses are $\rm 9.7 \times 10^6 M_{\odot}$ and $\rm 1.8 \times 10^6 M_{\odot}$, respectively. The reference model is calibrated to reproduce a subset of galaxy observations at present-day, such as the stellar mass function. Dark matter halos are over-dense regions in the simulated volume identified with the Friends-of-Friends\cite{1985ApJ...292..371D} and \textsc{subfind} \cite{2009MNRAS.399..497D} algorithms. Halo masses ($M_{200}$) correspond to all the matter within the radius $R_{200}$. The \textsc{eagle} database \cite{2016A&C....15...72M} provides information about integrated properties of these dark matter halos and their hosted galaxies.

\newpage 

\section*{Data Availability}
Ancillary data of the galaxy properties used in this study is available:  Stellar masses, star formation rates, apparent angular momentum, and the morphological classification can be found in table B.1 of ref. \cite{2019A&A...632A..59F}, while the stellar population properties can be found at https://cdsarc.cds.unistra.fr/viz-bin/cat/J/A+A/581/A103.

Total dynamical masses used in this study will be published in a forthcoming paper (Lyubenova et al. \textit{in prep.}), although they can be provided upon reasonable request contacting  \textit{mlyubeno@eso.org}. 

The group and cluster catalog from ref. \cite{2007ApJ...671..153Y} is available at https://gax.sjtu.edu.cn/data/Group.html. The \textsc{eagle} database \cite{2016A&C....15...72M} is public and it is available at https://icc.dur.ac.uk/Eagle/database.php

\section*{Code Availability}
We obtain the total dynamical masses of the galaxies by fitting axisymmetric dynamical models to their stellar kinematic
maps. Our dynamical models and codes used for their determination are described in detail in a forthcoming paper (Lyubenova et al. \textit{in prep.}).

\section*{Ethics and inclusion}
Our research project was conducted with integrity, respect, and inclusivity towards all members of the scientific community. We strive to create a safe, inclusive and supportive environment where everyone can contribute with their perspectives and ideas, regardless of their background or identity.

\section*{Acknowledgements}
We thank the referees for their constructive reports, their feedback and comments provided during this process. LSD thanks Sandra Faber, Joel Primack, David Koo, Joanna Woo, Doug Hellinger, Marc Huertas-Company, Farhanul Hasan and Joop Schaye for their insightful discussions. LSD, IMN and JFB acknowledge support through the RAVET project by the grant PID2019-107427GB-C32 from the Spanish Ministry of Science, Innovation and Universities (MCIU), and through the IAC project TRACES which is partially supported through the state budget and the regional budget of the Consejer\'ia de Econom\'ia, Industria, Comercio y Conocimiento of the Canary Islands Autonomous Community. IMN also acknowledges support from grant ProID2021010080 in the framework of Proyectos de I+D por organismos de investigaci\'on y empresas en las \'areas prioritarias de la estrategia de especializaci\'on inteligente de Canarias (RIS-3). FEDER Canarias 2014-2020.  ML acknowledges the ESO Science Support Discretionary Fund. GvdV acknowledges funding from the European Research Council (ERC) under the European Union’s Horizon 2020 research and innovation program under grant agreement No 724857 (Consolidator Grant ArcheoDyn). This research made use of Astropy, a community-developed core Python package for Astronomy \cite{astropy:2013,astropy:2018}, 
 and of the Numpy \cite{2020Natur.585..357H}, Scipy \cite{2020SciPy-NMeth}, Matplotlib \cite{2007CSE.....9...90H}, Emcee \cite{2013PASP..125..306F} and Pandas \cite{mckinney-proc-scipy-2010} libraries.

This version of the article has been accepted after peer
review, but is not the Version of Record and does not reflect post-acceptance improvements. The Version of Record is available online at:
http://dx.doi.org/10.1038/s41550-024-02209-8. Use of this Accepted Version is subject to the publisher's Accepted
Manuscript terms of use https://www.springernature.com/gp/open-research/policies/acceptedmanuscript-terms.

\section*{Author contributions statement}
LSD carried out the data analysis and wrote the paper. LSD, IMN and JFB conceived and desgined the project. ML performed the dynamical modelling and provided the dynamical mass measurements. GvdV was involved in the dynamical modelling. All authors interpreted and discussed the results, and commented on the manuscript.

\section*{Competing interests statement}
We declare that none of the authors have competing financial or non-financial interests as defined by Nature Portfolio.


\end{document}